\documentclass[english,aps,prl,twocolumn,superscriptaddress,showpacs,reprint]{revtex4}
\usepackage{lmodern}
\usepackage[T1]{fontenc}
\usepackage{xcolor}
\usepackage{pdfcolmk}
\usepackage{babel}
\usepackage{textcomp}
\usepackage{graphicx}
\PassOptionsToPackage{normalem}{ulem}
\usepackage{ulem}
\usepackage{subscript}
\usepackage[unicode=true,
 bookmarks=true,bookmarksnumbered=false,bookmarksopen=false,
 breaklinks=true,pdfborder={0 0 0},backref=false,colorlinks=true]
 {hyperref}

\makeatletter

\providecolor{lyxadded}{rgb}{0,0,1}
\providecolor{lyxdeleted}{rgb}{1,0,0}
\@ifundefined{textcolor}{}
{%
 \definecolor{BLACK}{gray}{0}
 \definecolor{WHITE}{gray}{1}
 \definecolor{RED}{rgb}{1,0,0}
 \definecolor{GREEN}{rgb}{0,1,0}
 \definecolor{BLUE}{rgb}{0,0,1}
 \definecolor{CYAN}{cmyk}{1,0,0,0}
 \definecolor{MAGENTA}{cmyk}{0,1,0,0}
 \definecolor{YELLOW}{cmyk}{0,0,1,0}
}

\makeatother

\begin{document}

\title{Inducing Lifshitz transition by extrinsic doping of surface bands
in topological crystalline insulator Pb\textsubscript{1-x}Sn\textsubscript{x}Se}

\author{I. Pletikosi\'{c}}

\email{ivop@princeton.edu}

\selectlanguage{english}%

\affiliation{Department of Physics, Princeton University, Princeton, NJ 08544,
USA}

\affiliation{Condensed Matter Physics and Materials Science Department, Brookhaven
National Laboratory, Upton, NY 11973, USA}

\author{G. D. Gu}

\author{T. Valla}

\affiliation{Condensed Matter Physics and Materials Science Department, Brookhaven
National Laboratory, Upton, NY 11973, USA}
\begin{abstract}
Narrow gap semiconductor Pb\textsubscript{1-x}Sn\textsubscript{x}Se
was investigated for topologically protected surface states in its
rock-salt structural phase for x=0.45, 0.23, 0.15, and 0. Angle-resolved
photoelectron spectroscopy of intrinsically \emph{p}-doped samples
showed clear indication of two Dirac cones, eccentric about the time-reversal
invariant point $\bar{\mathrm{X}}$ of the surface Brillouin zone
for all but the x=0 sample. Adsorption of alkalies gradually filled
the surface bands with electrons, driving the x>0 topological crystalline
insulator systems through Lifshitz transitions, and from a hole- to
electron-like Fermi surface. The electron-doped bands in x>0 samples
exhibited the full configuration of the Dirac cones, also confirming
electron-hole symmetry of the surface bands.
\end{abstract}


\maketitle
Symmetries of physical systems govern many of their fundamental properties.
In the case of topological insulators (TI), time-reversal symmetry
provides protection of gapless surface states accommodated in the
 bulk band gap \citep{Murakami2004,Kane2005}. These states appear
as an odd number of Dirac cones crossing the Fermi level, and are
known for their momentum-locked spin configuration, robustness against
weak disorder, absence of elastic backscattering, and a myriad of
exotic phenomena when interfaced with magnetism or superconductivity
\citep{Hasan2010,Qi2011}.

A related class of materials, topological crystalline insulators (TCI),
owe the protection of their gapless surface bands to crystal inversion
symmetries \citep{Fu2011}. The prediction of such states came from
symmetry considerations on a family of IV-VI semiconductors with strong
spin-orbit coupling, Pb\textsubscript{1-x}Sn\textsubscript{x}(Te,Se)
\citep{Hsieh2012,Volkov1985}, long known for the inversion of the
bulk band structure at two extremal compositions $x$ \citep{Strauss1967,Dalven1969,Calawa1969}.
Even though the surface states of these semiconductors become topologically
nontrivial  only when uniaxial deformation is introduced \citep{Fu2007a},
it was found that the mirror symmetry with respect to the {[}110{]}
plane gives necessary protection to spin-momentum locked states localized
at their (001), (111) or (110) surfaces. Surface states at (001),
in which the rock-salt crystals naturally cleave, comprise of four
Dirac cones centered slightly off the surface Brillouin zone boundary,
near the $\bar{\mathrm{X}}$ point ($k_{x|y}=0.73\,\AA^{-1}$, $k_{y|x}=0$)
on the mirror-symmetric line $\bar{\mathrm{\Gamma}}$--$\bar{\mathrm{X}}$--$\bar{\mathrm{\Gamma}}$.
These states have been observed in a recent series of angle-resolved
photoemission (ARPES) studies \citep{Xu2012a,Tanaka2012,Dziawa2012a,Tanaka2013a,Wojek2013}.
It was found that (001) surface states of Pb\textsubscript{1-x}Sn\textsubscript{x}Te
are topologically trivial \citep{Tanaka2013a} and gapped \citep{Xu2012a}
for $x<0.3$, but develop to two Dirac cones moving away from $\bar{\mathrm{X}}$
for $x>0.3$, all up to $x=1$ \citep{Tanaka2013a}. Also, clear manifestations
of TCI spin-locked states have been found for the (001) surface of
Pb\textsubscript{0.73}Sn\textsubscript{0.27}Se, as well as their
temperature dependent transition from topological to trivial \citep{Dziawa2012a,Wojek2013}.
All these lack a precise mapping of the low-energy configuration of
the surface bands, which is a crucial element in determining quasiparticle
scattering patterns and possible scattering protections \citep{Zeljkovic2013,Gyenis2013}.

A peculiar feature in the electronic structure of these states is
the existence of so called Lifshitz points, where the change in topology
of constant energy contours occurs. By varying the filling of the
surface states, the Lifshitz transitions could, in principle, be induced,
with interesting consequences on surface transport. This, and a wealth
of exotic features, from extremely high surface mobilities to ferroelectricity
and superconductivity, but also the possibility of their tuning by
strain, doping or alloying, make these materials worth exploring both
from the fundamental perspective and for various applications. 

Here we investigate (001) surface states of Pb\textsubscript{1-x}Sn\textsubscript{x}Se
over a range of compositions for which the crystals grow in rock-salt
structure. We set the lower limit on $x$ at which the transition
from TCI to trivial surface states occurs, and give a precise experimental
characterization of the two interlocking Dirac cones. We also show
that initially \emph{p}-type surfaces can be gradually turned into
\emph{n}-type by adsorbing electron donors, demonstrating the possibility
of a transformation of the Fermi surface topology through the Lifshitz
transition. 

Single crystal ingots of Pb\textsubscript{1-x}Sn\textsubscript{x}Se
were grown in a custom built floating zone apparatus from the  stoichiometric
melt of high purity (5N) constituent elements. Samples of a few cubic
millimeters were cut from the ingots and cleaved \emph{in situ} for
the experiments\emph{.}

\begin{figure*}[th]
\includegraphics[width=170mm]{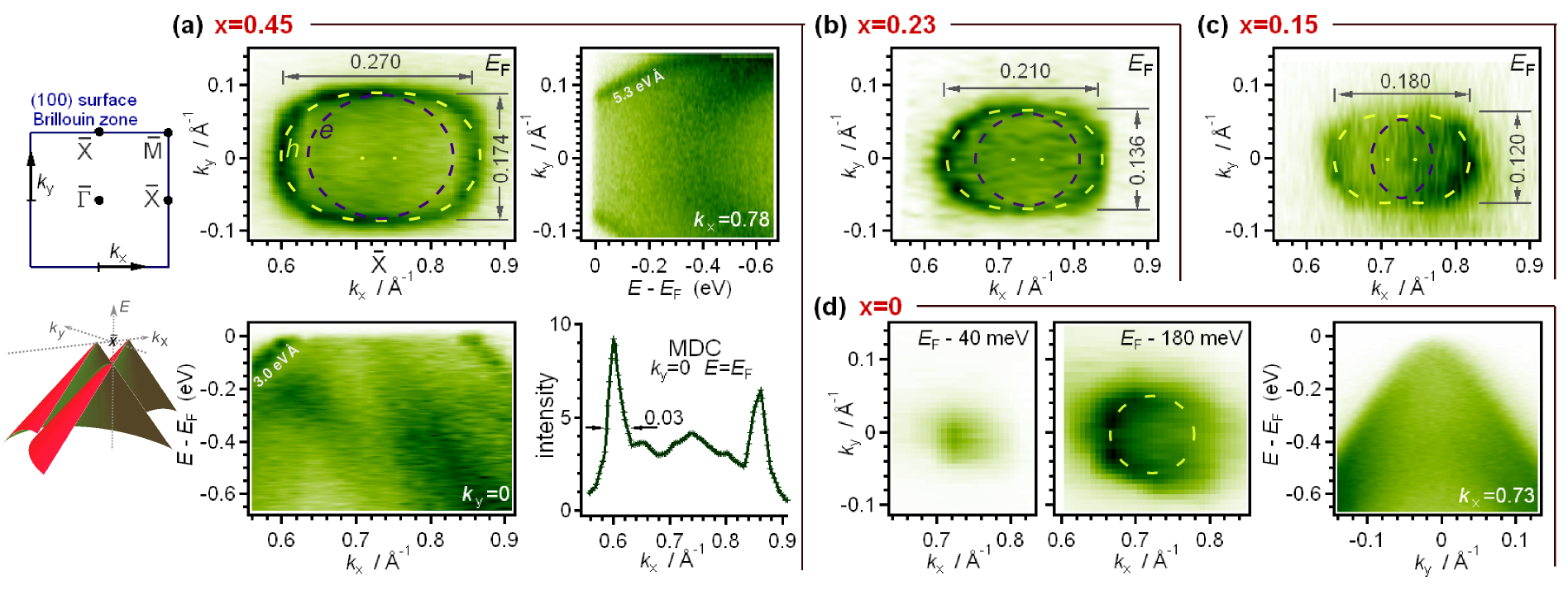}

\caption{\label{fig:intrinsic}(color online) ARPES intensity maps taken at
18~K from the (001) surface of intrinsically p-doped Pb\textsubscript{1-x}Sn\textsubscript{x}Se
samples in the vicinity of the $\bar{\mathrm{X}}$ point of the surface
Brillouin zone ($k=0.73\,\AA^{-1}$) for four different compositions
$x$. (a) Fermi surface, band dispersions along $k_{y}$ (at $k_{x}=0.78\,\AA^{-1}$)
and $k_{x}$(at $k_{y}=0$), and momentum distribution curve (MDC)
of the bands close to the Fermi level ($E_{\mathrm{F}}$) along the
$\bar{\mathrm{\Gamma}}$--$\bar{\mathrm{X}}$--$\bar{\mathrm{\Gamma}}$
direction for $x=0.45$; (b) and (c) constant energy maps for $x=0.23$
and $x=0.15$ samples at the Fermi energy; (d) constant energy maps
for the $x=0$ sample 40 and 180~meV below the Fermi level, and the
band dispersion map along $k_{y}$ at $\bar{\mathrm{X}}$. Dashed
lines added to the constant energy maps represent assumed band configuration
of two Dirac cones slightly eccentric about $\bar{\mathrm{X}}$ (calculated
by a formula from Ref. \onlinecite{Liu2013a}). Electron/e and hole/h
pockets are separately marked.}
\end{figure*}
 ARPES measurements have been conducted at the U13 beamline ($h\nu$
from 17 to 22 eV) of the National Synchrotron Light Source at Brookhaven
National Laboratory using a Scienta 2002 analyzer with an overall
energy resolution of 15 meV and angular resolution <0.2\textdegree{}
. Brillouin zone mapping was accomplished by polar angle rotation,
perpendicularly to the analyzer slit, in steps of 0.5\textdegree{}.
Liquid helium was used to cool the samples down to 10 K.

ARPES intensity maps for $x>0$, Figs. \ref{fig:intrinsic}(a)--(c),
show a region of higher intensity centered at $\bar{\mathrm{X}}$,
encircled by hole-like bands, and low background intensity beyond.
Unlike the inner, broad and mostly featureless bands, the outer bands
are sharp, having the Fermi level momentum distribution curve (MDC)
width at half maximum of $0.03\,\AA^{-1}$ . These bands show no dependence
on the excitation energy, which by not having out-of-surface momentum
$k_{z}$ as a quantum number, signifies their surface localization
(cf. Fig. S1 in Supplemental Material \citep{SupplemMat}).  Band
dispersion spectra, like the two shown in Fig. 1(a) taken along and
perpendicular to the $\bar{\mathrm{\Gamma}}$--$\bar{\mathrm{X}}$--$\bar{\mathrm{\Gamma}}$
direction, exhibit linear bands in all directions over the energy
region probed. 

The elongated shape of the constant energy cuts of the outer (surface)
states points to their origin in two interlocking Dirac cones slightly
displaced from $\bar{\mathrm{X}}$ along the $\bar{\mathrm{\Gamma}}$--$\bar{\mathrm{X}}$--$\bar{\mathrm{\Gamma}}$
direction, as already found for Pb\textsubscript{0.77}Sn\textsubscript{0.23}Se
\citep{Dziawa2012a} or Pb\textsubscript{1-x}Sn\textsubscript{x}Te
\citep{Tanaka2012,Xu2012a,Tanaka2013a}. And indeed, the $k_{x}$\,--\,$k_{y}$
constant energy maps, as those shown in Fig. \ref{fig:intrinsic}
for the Fermi energy, are easily fitted by the contours of a model
band dispersion given by Liu \emph{et al}. \citep{Liu2013a}, with
the parameters adjusted in order to reproduce the observed band velocities
($v_{x}$, $v_{y}$) and the overall shape governed by ($m$, $\delta$)
which parametrize intervalley scattering at the lattice scale. Both
sets of parameters greatly differ from those that Liu \emph{et al}.
obtained by fitting to density-functional \emph{ab initio} bands (Figure
3 in \citet{Liu2013a}, $m=70$, $\delta=26$, $v_{x}=2.4$, $v_{y}=1.3$).
While our values of $m=6\pm3\,\mathrm{meV}$, $\delta=68\pm3\,\mathrm{meV}$
and $v_{y}=2.8\pm0.3\,\mathrm{eV}\AA$ (in the geometry used by Liu
\emph{et al}.) appear to be universal for all the fitted bands, velocity
$v_{x}$ in the direction perpendicular to $\bar{\mathrm{\Gamma}}$--$\bar{\mathrm{X}}$--$\bar{\mathrm{\Gamma}}$
had to be tuned from $5.3\,\mathrm{eV}\AA$ for the $x=0.45$ sample,
to $4.4\,\mathrm{eV}\AA$ for $x=0.23$, and to $3.5\,\mathrm{eV}\AA$
for the $x=0.15$ sample. This variation probably indicates that the
model is only applicable to the parts of the bands close to the Dirac
points, and that considerable warping takes place at energies farther
away. As we will see below, our values for $m$ and $\delta$ result
in highly circular constant energy contours, in contrast to distorted
contours obtained in most theoretical considerations \citep{Safaei2013,Liu2013a,Fang2013}.

\begin{figure}[th]
\includegraphics[width=78mm]{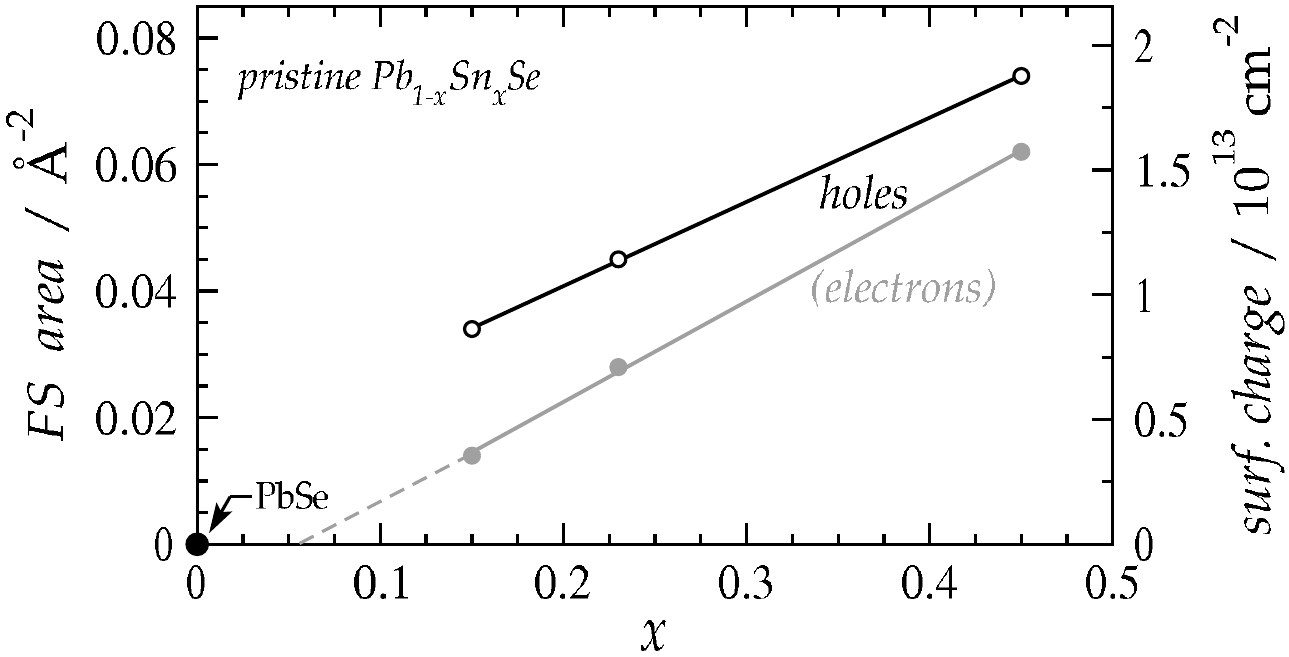}

\caption{\label{fig:FSarea}Tin content $x$ dependence of hole and electron
pocket Fermi surface areas for surface states in intrinsic \emph{p}
type Pb\textsubscript{1-x}Sn\textsubscript{x}Se. }
\end{figure}
From Figs. \ref{fig:intrinsic}(a)--(c) it is obvious that the Fermi
surfaces (FS) of the $x>0$ samples shrink in size with decreasing
Sn content $x$. Taking the measures of the contours, we calculated
the FS areas for the hole and electron pockets in the first surface
Brillouin zone and plotted them in Fig. \ref{fig:FSarea}. The dependence
on $x$ is manifestly linear. A small variation among the samples
cut from the same ingot was found, as the intrinsic doping (thus the
Fermi surface size) is known to be dependent on the growth conditions,
defects present, etc. \citep{Dziawa2012a,Melngailis1972,Dixon1971,Dalven1969}
As we will explain below, only the hole charge contained in surface
bands(right axis in Fig. \ref{fig:FSarea}), has physical relevance
in our samples, the charge from the electron pockets being delocalized
in the bulk. Extrapolating the derived electron pocket FS size to
zero, the composition of $x\approx0.05$ would correspond to a point
of Lifshitz transition in the pristine sample. Before reaching that
point, however, a subtle quantum phase transition might occur that
would make the bands topologically trivial.

Comparison of the spectra for $x>0$ in Figs. \ref{fig:intrinsic}(a)--(c)
with the spectra for PbSe ($x=0)$ in Fig. \ref{fig:intrinsic}(d),
where a single cone isotropic around X is found, leaves  no doubt
that all three $x>0$ samples are topological crystalline insulators.
This has been further elucidated by extrinsic doping of the bands,
which unveiled the upper Dirac cones and also made all their branches
visible.

\begin{figure}[th]
\includegraphics[width=85mm]{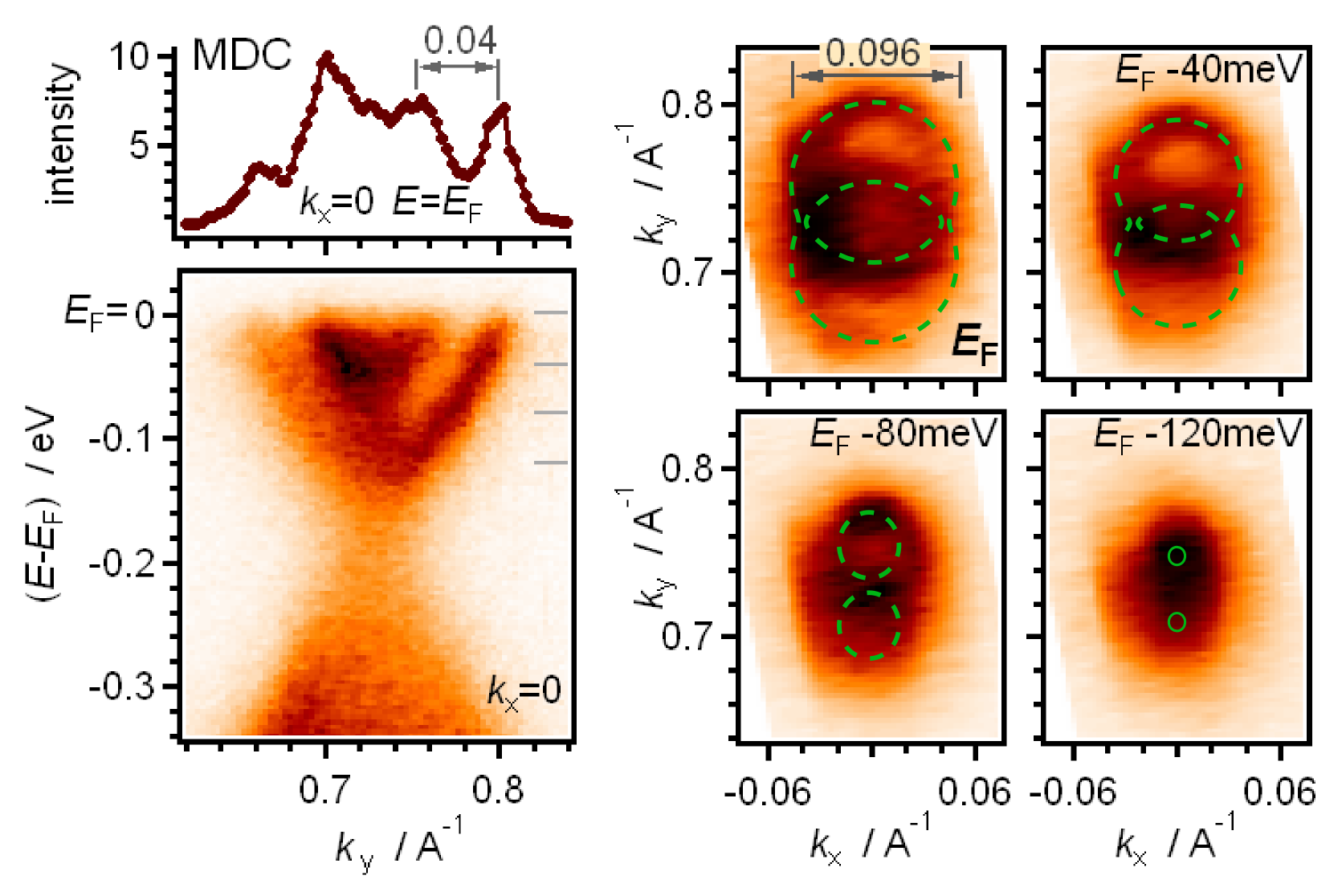}

\caption{\label{fig:doped}(color online) ARPES intensity maps from the electron
doped (001) surface of Pb\textsubscript{0.85}Sn\textsubscript{0.15}Se
($x=0.15$). The doping was achieved by rubidium adsorption at 18~K.
Momentum distribution curve (MDC) is from the band dispersion spectrum
taken along the $k_{y}$ direction in the vicinity of $\bar{\mathrm{X}}$
($k_{x}=0$). Constant energy cuts of the two Dirac cones eccentric
about $\bar{\mathrm{X}}$, made at the Fermi level ($E_{\mathrm{F}}$)
and 40, 80, and 120~meV below are shown on the right.}
\end{figure}

It is well known that alkali atoms when adsorbed onto a surface give
away their electrons, the charge transfer to the substrate being dependent
on the coverage. This has been demonstrated as a very effective way
of electron-doping of surfaces of TIs and tuning of the band filling,
similar to the gate-doping in transport devices \citep{Valla2012}.
 Here, upon a gradual lowering of the electronic bands, saturation
appears and no further doping is possible. Given its low intrinsic
\emph{p}-doping, the $x=0.15$ sample was easiest to dope with electrons,
exposing most of the electronic structure above the Dirac point. A
band dispersion ARPES spectrum on the left of Fig.  \ref{fig:doped}
shows a cut through the electronic structure of a Rb covered $x=0.15$
sample along the $\bar{\mathrm{\Gamma}}$--$\bar{\mathrm{X}}$--$\bar{\mathrm{\Gamma}}$
high-symmetry line. For the sake of a precise momentum-space mapping,
it was taken at an elevated tilt angle---along the $k_{y}$ direction.
The spectrum unambiguously presents a cross-section of two Dirac cones
shifted by $0.04\,\AA^{-1}$ along the $\bar{\mathrm{\Gamma}}$--$\bar{\mathrm{X}}$
direction in momentum space, one cone showing brighter than the other.
Their apex is found at 0.14~eV below the Fermi level. Compared to
the $0.16$--$0.28\,\AA^{-1}$ cone separation in Pb\textsubscript{1-x}Sn\textsubscript{x}Te
\citep{Tanaka2013a,Xu2012a}, the separation of $0.04\,\AA^{-1}$
in Pb\textsubscript{1-x}Sn\textsubscript{x}Se makes the analysis
of the tangled bands more difficult. Only due to relatively sharp
surface bands, having the width at half maximum as low as $0.022\AA^{-1}$,
we were able to determine the details of their configuration with
sufficient precision. Constant energy maps of the region of momentum
space around $\bar{\mathrm{X}}$ show a pattern of two circles always
centered about the same points. Slices of the cones at four different
energies referenced to the Fermi level, displayed on the right side
of Fig. \ref{fig:doped}, have been fitted by the contours of a theoretical
model by Liu \emph{et al}. \citep{Liu2013a} (using $m=6\,\mathrm{meV}$,
$\delta=67\,\mathrm{meV}$, and $v_{x}=2.9\,\mathrm{eV}\AA$, $v_{y}=3.1\,\mathrm{eV}\AA$).
Note that these parameters are very similar to those needed to fit
the pristine bands, suggesting that the surface state is particle-hole
symmetric. The contours start at the Dirac point, roughly 140~meV
below the Fermi level, like two separate circular electron pockets,
but then go to a topologically distinct configuration around -80~meV,
in which the outer contour forms an electron pocket while the inner,
interlocked part, forms a hole pocket. The transition between the
two configurations is easy to induce, as the electron doping by alkali
deposition can continuously sweep the chemical potential through the
portions of the lower and upper Dirac cones.

\begin{figure}[th]
\includegraphics[width=80mm]{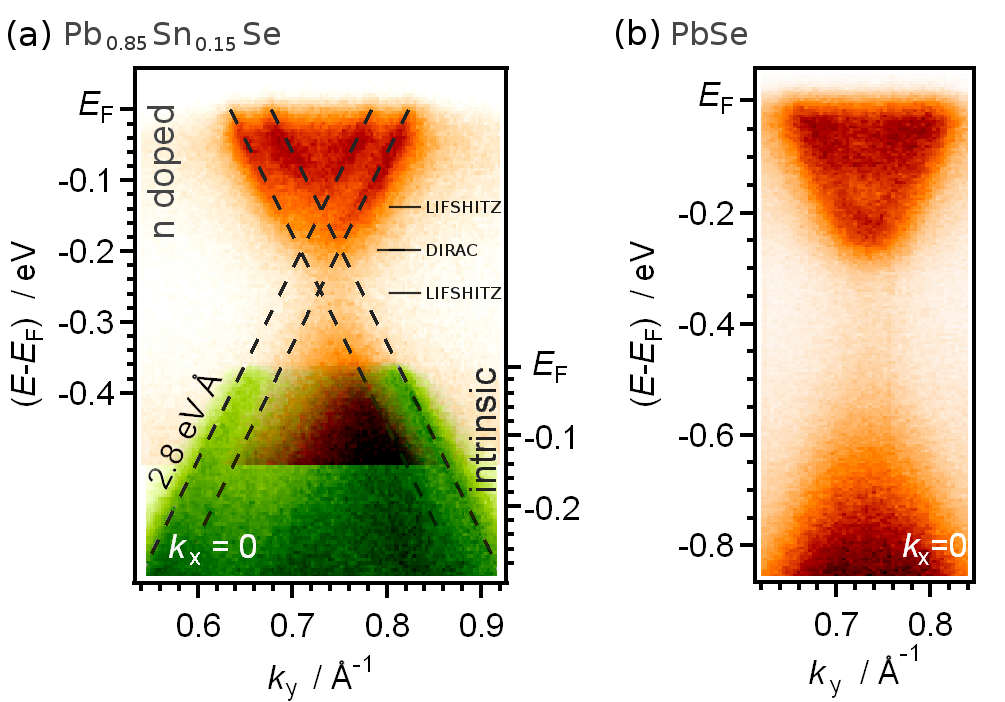} 

\caption{\label{fig:superposition}(color online) (a) ARPES spectrum showing
highly \emph{n}-doped surface bands in Pb\textsubscript{0.85}Sn\textsubscript{0.15}Se
($x=0.15$) realized by rubidium adsorption at 10~K (orange) is superimposed
onto the spectrum of the pristine, intrinsically \emph{p}-doped sample
(green). Dashed lines match the four bands of the upper Dirac cones
and two bands of the lower cones. (b) ARPES spectrum of highly doped
bands of PbSe. All spectra were taken along the $\bar{\mathrm{\Gamma}}$--$\bar{\mathrm{X}}$--$\bar{\mathrm{\Gamma}}$
direction. Note different energy scales in (a) and (b). }
\end{figure}

To establish the relation between the upper and lower bands,  we
superimpose in Fig. \ref{fig:superposition}(a) the band dispersion
images taken from the intrinsically \emph{p}-doped $x=0.15$ sample
along the $\bar{\mathrm{\Gamma}}$--$\bar{\mathrm{X}}$--$\bar{\mathrm{\Gamma}}$
direction, and the same sample in the highest \emph{n}-doping achieved
by rubidium adsorption at 10~K.  Having the two lower bands extended
by dashed lines, a position can be found where they agree with two
bands of the upper cones.  An extension of the remaining two upper
bands to lower energies shows their coincidence with the area of increased
intensity that has been ascribed to photoemission from the bulk. 
By having a possibility of bulk propagation, these states do not show
up as distinct surface bands. This is why the inner, interlocked bands
traced by the theoretical contours in Fig. \ref{fig:intrinsic} were
not resolved on the Fermi surface maps of any of the $x>0$ samples.
The same applies to the lower bands in doped samples: as the energy
shift by doping of the surface and bulk bands is not the same, these
states probably overlap with the bulk states, therefore losing their
surface localization. The upper Dirac cones of the doped samples are
outside of the region of the bulk bands, and are all clearly resolved. 

The overlay of the two spectra makes possible assessing their doping
levels with respect to the Dirac point: the pristine sample exhibits
a 170~meV \emph{p}-doping, while the sample of the highest extrinsic
doping shows\emph{ }surface bands\emph{ }lowered by 190~meV. We have
thus made a transition from a hole-like Fermi surface of the pristine
$x=0.15$ sample, to a similar size electron-like  Fermi surface
by \emph{in-situ} doping. On the way, surface bands went through two
Lifshitz transitions, 67~meV on either side of the Dirac crossing.
 Similar position of the Lifshitz transition can be inferred from
a tunneling study by Okada \emph{et al}. \citep{Okada2013a} They
interpret the maximum in $\mathrm{d}I/\mathrm{d}V$ curves $\pm40\,\mathrm{meV}$
from the Dirac point as the position of a van Hove singularity in
the density of states. This places the Lifshitz points at $\pm70\,\mathrm{meV}$
($E_{DP2\pm}$ in Fig. 1D, Ref. \onlinecite{Okada2013a}). 

We note that the two outer bands of highly doped Pb\textsubscript{0.85}Sn\textsubscript{0.15}Se
samples (Figs. \ref{fig:doped} and \ref{fig:superposition}; also
raw spectra in Supplemental Material \citep{SupplemMat}, Fig. S2)
show an abrupt change of velocity, approximately at the level of the
Lifshitz crossing. This has not been captured by the $k\cdot p$ formula
by Liu \emph{et al.} \citep{Liu2013a} which gives strikingly linear
bands, nor a more elaborate Hamiltonian by Fang \emph{et al.} \citep{Fang2013}
was able to reproduce the kink in our previous work, Ref. \onlinecite{Gyenis2013},
tending to bend the bands in the direction opposite of the observed.
One should therefore consider if the change in the band topology at
the Lifshitz transition or the nearby van Hove singularity in the
density of states coming from the band minimum along $\bar{\mathrm{X}}$--$\bar{\mathrm{M}}$
marks an onset of an interaction that would renormalize the bands.
This, as well as structural mirror-symmetry breaking proposed by Okada
\emph{et al}. \citep{Okada2013a}, could lead to mass acquisition
in the Dirac bands.

In a final test, the $x=0$ sample was electron doped by rubidium
adsorption. As Fig.  \ref{fig:superposition}(b) shows, no double-cone
structure was found, but a single upper surface state with a curved
apex, separated from the lower states by a gap of at least 200~meV.
Similar gapped states were found in topologically trivial Pb\textsubscript{0.8}Sn\textsubscript{0.2}Te
\citep{Xu2012a}. Interestingly enough, the $x=0$ case appears to
be a limiting case of the two spin-locked Dirac cones coalescing into
one as the Sn content is being reduced. Trends are clear: the cones
in Pb\textsubscript{1-x}Sn\textsubscript{x}Te were found approaching
each other from $0.26\,\AA^{-1}$ at x=0.5 to $0.16\,\AA^{-1}$ at
x=0.3 \citep{Tanaka2013a}; in Pb\textsubscript{1-x}Sn\textsubscript{x}Se,
a separation of $0.054\,\AA^{-1}$  found for x=0.23 \citep{Dziawa2012a}
reduces to $0.04\,\AA^{-1}$ for x=0.15. It is interesting to note
that the surface doping does not affect the eccentricities or any
other aspect of the surface electronic structure, implying that these
properties are determined by the bulk and that the extrinsic doping
does not break any relevant crystal symmetries.

To summarize, (001) surface states of Pb\textsubscript{1-x}Sn\textsubscript{x}Se
appeared as two Dirac cones eccentric about the $\bar{X}$ point of
the surface Brillouin zone for x=0.45, 0.23, and 0.15, but a single
cone for x=0, indicating a transition from topologically protected
to trivial states at Sn content lower than 0.15, in contrast to Pb\textsubscript{1-x}Sn\textsubscript{x}Te
system which ceases to be a TCI already at x=0.3 \citep{Tanaka2013a}.
Electron doping by rubidium adsorption drove the system through Lifshitz
transitions and showed that, aside from the differences in the overlap
with the bulk bands, the surface states are particle-hole symmetric.

\emph{}
\begin{acknowledgments}
This work was supported by ARO MURI program, grant W911NF-12-1-0461,
and US Department of Energy, Office of Basic Energy Sciences, contract
no. DE-AC02-98CH10886. 
\end{acknowledgments}

\end{document}